\documentclass[twocolumn,prl,floatfix,showpacs]{revtex4}
\usepackage{graphicx}
\usepackage{dcolumn}
\usepackage{amsmath}
\usepackage{amssymb}

\begin{document}

\title{Polynomial-Time Simulation of Pairing Models on a Quantum Computer}
\author{L.-A. Wu} 
\author{M.S. Byrd} 
\altaffiliation[Present address: ]{Harvard University, Maxwell Dworkin Laboratory, 33 Oxford Street, Cambridge, Massachusetts 02138}
\author{D.A. Lidar}
\affiliation{Chemical Physics Theory Group, University of Toronto, 80 St. George St.,
Toronto, Ontario M5S 3H6, Canada}

\begin{abstract}
We propose a polynomial-time algorithm for simulation of the class of pairing
Hamiltonians, e.g., the BCS Hamiltonian, on an NMR quantum computer. The
algorithm adiabatically finds the low-lying spectrum in the vicinity of the
gap between ground and first excited states, and provides a test of the
applicability of the BCS Hamiltonian to mesoscopic superconducting systems,
such as ultra-small metallic grains.
\end{abstract}

\maketitle

The potential of quantum computers (QCs) to provide exponential
speed-up in the simulation of quantum physics problems was originally
conjectured by Feynman \cite{Feynman:QC}, confirmed by Lloyd
\cite{Lloyd:96}, and later studied theoretically by a number of
authors, e.g., \cite{Meyer:96Wiesner:96Zalka:98Boghosian:97aLidar:98RC,Abrams:99,Abrams:97Ortiz:00,Bravyi:00,Dodd:01Wocjan:01}. NMR-QC
experiments performing quantum physics simulations were reported in
\cite{Somaroo:99Tseng:00}. Current QC technology is limited to fewer
than 10 qubits and the testing of simple algorithms \cite{Knill:00a}.
QCs of the next generation, with 10-100 qubits, have the potential to
solve hard problems in quantum many-body theory. We show here how this
observation can be applied to the problem of simulating the class of
\emph{pairing} Hamiltonians with general, i.e., \emph{arbitrary
long-range} interactions. The pairing Hamiltonians are of wide
interest in condensed matter and nuclear physics
\cite{Mahan:bookRing:book}. An important example of a pairing
Hamiltonian is the BCS model of low-T$_{\mathrm{c}}$
superconductivity. We provide an algorithm for testing the validity of
the \emph{general} BCS Hamiltonians of finite particle-number systems,
pertinent to nuclear systems and mesoscopic condensed-phase systems,
such as ultra-small metallic grains
\cite{Ralph:97,Mastellone:98,BraunDelft:98,Dukelsky:99}. These grains
provide a fertile testing ground for the BCS ansatz for the ground
state wave function. The BCS wave function is a superposition of
different Fermion numbers and is expected to be exact in the
thermodynamic limit \cite{Anderson:58}. In contrast, in ultra-small
metallic grains the number of states $N$ within the Debye frequency
cutoff from the Fermi energy is only $ \sim 100$. A similar estimate
holds for the number of states within a few major shells for medium
or heavy nuclei. In systems with finite particle number the BCS ansatz
is doubtful, and at the same time exact numerical diagonalization of
the \emph{general} BCS Hamiltonian is impractical beyond a few tens of
electron pairs \cite{Mastellone:98}. Various approximations have been
proposed \cite{Braun:98}, but it would clearly be desirable to have an
exact numerical solution for the problem. In
\cite{Abrams:97Ortiz:00,Bravyi:00} efficient QC algorithms were
presented 
for simulating a many-body fermionic system. While the BCS Hamiltonian
describes a system of interacting fermions, it does so at the level of
an effective field theory. This can be expressed in terms of an
interacting spin system \cite{Anderson:58}, or parafermions
\cite{WuLidar1:01}. Therefore the fermionic simulation algorithms
\cite{Abrams:97Ortiz:00} are not directly applicable. Further, while a
number of authors have recently considered simulation of one
Hamiltonian in terms of another \cite {Dodd:01Wocjan:01}, the
connection of these phenomenological Hamiltonians to those of
many-body condensed matter and nuclear physics is not \textit{a
priori} clear. Here we clarify the correspondence by proposing an
explicit and numerically exact diagonalization algorithm that is
suitable for general pairing Hamiltonians, and is \emph{directly
implementable} in NMR-type quantum computers \cite{Cory:00}. More
generally, with minor modifications our algorithm is applicable to all
QCs with short-range exchange-type interactions, such as quantum dots
\cite{Burkard:00}.
Using an adiabatic procedure, we show how to obtain
only the low-lying energy spectrum, e.g., in the vicinity of the
superconducting gap, with an algorithm that takes
$\sim N^4$, instead of exponential, computational steps. The number of qubits we require equals the
effective number of states $N$, so that a QC with $\sim 100$ qubits
(neglecting overhead due to error correction) could solve a problem
that is well out of the reach of current classical computers.

\textit{Mapping of Bosons and Fermions to Qubits.---} Pairing Hamiltonians
are typically expressed in terms of fermionic or bosonic creation
(annihilation) operators, $c_{m}^{\dagger }$ ($c_{m}$) and $b_{m}^{\dagger }$
($b_{m}$), respectively, where $|m|=1,2,\ldots ,N$ denotes all relevant
quantum numbers. E.g., the general BCS pairing Hamiltonian has the form: 
\begin{equation*}
H_{\mathrm{BCS}}=\sum_{m=1}^{N}\frac{\epsilon _{m}}{2}
(n_{m}^{F}+n_{-m}^{F})+\sum_{m,l=1}^{N}V_{ml}^{+}c_{m}^{\dagger
}c_{-m}^{\dagger }c_{-l}c_{l}
\end{equation*}
where $n_{\pm m}^{F}\equiv c_{\pm m}^{\dagger }c_{\pm m}$ is the number
operator, and the matrix elements $V_{ml}^{+}\equiv \left\langle
m,-m\right\vert V\left\vert l,-l\right\rangle $ (we impose no restriction on 
$m,l$) are real and can be calculated, e.g., for superconductors, in terms
of the Coulomb force and the electron-phonon interaction \cite{Mahan:bookRing:book}. Pairs of fermions are labeled by the quantum
numbers $m$ and $-m$, according to the Cooper pair situation where paired
electrons have equal energies but opposite momenta and spins: $m=(\mathbf{p},\uparrow )$ and $-m=(-\mathbf{p},\downarrow )$. These are degenerate, 
time-reversed partners
whose energies are considered phenomenological parameters \cite{Braun:98}.
The same idea is applicable to nuclei, where effective pairings occur
between nucleons in time-reversed partners \cite{Mahan:bookRing:book}. $N$
is an effective state number, which equals the number of qubits in the
algorithm below. E.g., in the case of metallic grains $N$ is
twice the the Debye frequency in units of the average level spacing
(inversely proportional to volume of the grain). For nuclear pairing models, 
$N$ could be the number of states in one or more major energy shells.

To make a connection to quantum algorithms we map the fermionic or bosonic
operators to qubit operators. We denote the raising and lowering operators
for the $m^{\mathrm{th}}$ qubit by the Pauli matrices $\sigma _{m}^{\pm }$,
acting non-trivially only on the $m^{\mathrm{th}}$ qubit. A
\textquotedblleft number operator\textquotedblright\ is $n_{m}=(\sigma_{m}^{z}+1)/2$, where $n_{m}=1$ ($0$) if the $m^{\mathrm{th}}$ qubit is in
state $|1\rangle $ ($|0\rangle $); $n=\sum_{m}n_{m}$ is the number of 1's in
a computational basis state (a ket of a single bit-string), and will
correspond, e.g., to the number of Cooper pairs in our applications below.
The computational ground state $\left\vert \mathtt{0}\right\rangle
=\left\vert 0_{1}0_{2}\cdots 0_{N}\right\rangle $ acts as a vacuum state: $
n\left\vert \mathtt{0}\right\rangle =\sigma _{m}^{-}\left\vert \mathtt{0}
\right\rangle =0$. Now we can consider three generic pairing cases and map
them to qubits. In each case we identify fermionic or bosonic operator pairs
that satisfy the commutation rules of $sl(2)=\{\sigma _{m}^{+},\sigma
_{m}^{-},\sigma _{m}^{z}\}$ (see \cite{WuLidar1:01} for details). These
cases are:\ (i) \emph{Fermionic particle-particle pairs} (e.g., Cooper
pairs): $sl(2)=\{c_{-m}c_{m},c_{m}^{\dagger }c_{-m}^{\dagger
},n_{m}^{F}+n_{-m}^{F}-1\}$, provided $n_{m}^{F}=n_{-m}^{F}$ (a condition
satisfied by $H_{\mathrm{BCS}}$), and $\left\vert \mathtt{0}\right\rangle
=\left\vert \mathtt{0}\right\rangle _{F}$. (ii)\emph{\ Fermionic
particle-hole pairs} (e.g., excitons):$\ sl(2)=\{c_{-m}^{\dagger
}c_{m},c_{m}^{\dagger }c_{-m},n_{m}^{F}-n_{-m}^{F}\}$, provided $
n_{m}^{F}+n_{-m}^{F}=1$ and $\left\vert \mathtt{0}\right\rangle
=c_{-N}^{\dagger }\cdots c_{-2}^{\dagger }c_{-1}^{\dagger }\left\vert 
\mathtt{0}\right\rangle _{F}$ . (iii) \emph{Bosonic `particle-hole' pairs}
(e.g., dual-rail photons in the optical quantum computer proposal \cite{Knill:00}): $sl(2)=\{b_{-m}^{\dagger }b_{m},b_{m}^{\dagger
}b_{-m},n_{m}^{B}-n_{-m}^{B}\}$, provided $n_{m}^{B}+n_{-m}^{B}=1$ and $
\left\vert \mathtt{0}\right\rangle =b_{-N}^{\dagger }\cdots b_{-2}^{\dagger
}b_{-1}^{\dagger }\left\vert \mathtt{0}\right\rangle _{B}$. The three
conditions above each restrict the dynamics to a different subspace of the
entire Hilbert space. The conditions play the role of conserved quantities
and only Hamiltonians that satisfy them preserve such subspaces.

It is now clear how to express $H_{\mathrm{BCS}}$ in terms of qubit
operators. In fact, a more general Hamiltonian, that is applicable to all
cases (i)-(iii) is: 
\begin{equation}
H_{\mathrm{p}}=\sum_{m=1}^{N}\frac{\varepsilon _{m}}{2}\sigma
_{m}^{z}+\sum_{r=\pm }\sum_{l>m=1}^{N}\frac{V_{ml}^{r}}{2}(\sigma
_{m}^{x}\sigma _{l}^{x}+r\sigma _{m}^{y}\sigma _{l}^{y}),  \label{eq:Hp}
\end{equation}
where $\varepsilon _{m}=\epsilon _{m}+V_{mm}^{+}$ and $V_{ml}^{-}=0$ for $H_{
\mathrm{BCS}}$; $l,m$ now denote both state indices and qubit indices.
Further, in the BCS case the qubit state space $\mathcal{H}_{P}=\mathrm{Span}
\{\left\vert \mathtt{0}\right\rangle ,\sigma _{m}^{+}\left\vert \mathtt{0}
\right\rangle ,\sigma _{l}^{+}\sigma _{m}^{+}\left\vert \mathtt{0}
\right\rangle ,\cdots \}$ is mapped into a subspace of the total fermionic
Hilbert space where $n_{m}^{F}=n_{-m}^{F}$. $H_{\mathrm{BCS}}$ conserves the
total number operator $n$ (the number of Cooper pairs). In terms of qubits,
this means that the number of $|1\rangle $'s in a general $N$-qubit state is
fixed by $H_{\mathrm{BCS}}$. Thus the Hilbert space splits into invariant
subspaces with dimension ${\ {\binom{N}{n}}}$ for fixed $n$. The
problem is reduced to diagonalizing separate blocks of size $
{\ {\binom{N}{n}}}$. For half-filled states in a system with $N=100$, an
exact solution could require diagonalizing a $
10^{29}\times 10^{29}$-dimensional matrix. Such a task is clearly
unfeasible on a classical computer.

\textit{Simulation of }$H_{\mathrm{p}}$.--- For concreteness and direct
contact with feasible experiments, we limit our discussion of the simulation
of $H_{\mathrm{p}}$ to the nearest-neighbor Ising-type Hamiltonian of NMR:\ $
H_{\mathrm{NMR}}=\sum_{l=1}^{N}\frac{\omega _{l}}{2}\sigma
_{l}^{z}+\sum_{l=1}^{N-1}J_{l}\sigma _{l}^{z}\sigma _{l+1}^{z}$,
supplemented with external single qubit operations $F=
\sum_{l=1}^{N}f_{l}^{x}\sigma _{l}^{x}+f_{l}^{y}\sigma _{l}^{y}$. The same
Hamiltonian describes, e.g., a QC implementation using coupled Josephson
junctions \cite{Blais:00}. We emphasize that this simulation is
also directly implementable in systems that use exchange-type interactions,
since the logical operations for those systems are
equivalent (up to polynomial overhead) to those using the Ising coupling 
\cite{Dodd:01Wocjan:01,WuLidar1:01}. We shall for simplicity
only explicitly discuss the case $V_{ml}^{-}=0$, but the same procedure will
apply also to the case of $V_{ml}^{-}\neq 0$ (since the two cases are
related by a simple unitary transformation). From now on we denote $
V_{ml}^{+}\equiv V_{ml}$.

Below, we develop an explicit polynomial-time algorithm for simulating $\{U_{
\mathrm{p}}(k\tau )=\exp (-iH_{\mathrm{p}}k\tau )\}_{k=1}^{T/\tau }$ ($\tau $, $T$ are defined later). This sequence can be Fourier-transformed and the
spectrum of $H_{\mathrm{p}}$ found \cite{Abrams:99}. However, although this
may be achieved directly using NMR methods, we are primarily interested in
the low-lying spectrum (e.g., in the BCS case, near the superconducting
gap). Our algorithm therefore includes an adiabatic component, that allows
us to probe just this part of the spectrum. Let us now outline the main
steps in our algorithm for simulating $H_{\mathrm{p}}$ using $H_{\mathrm{NMR}
}$ and $F$. (i) Prepare a computational basis state $|x_{n}\rangle $ with
fixed $n$ (number of $|1\rangle $'s). This step is well-known and
needs no further explanation \cite{Cory:00}. (ii) Quasi-adiabatically evolve $
|x_{n}\rangle $ to $|\psi (0)\rangle _{0}=|g_{n}\rangle +\theta
|e_{n}\rangle $: an approximate ground state of $H_{\mathrm{p}}$ ($
|g_{n}\rangle $ is an exact ground state, $|e_{n}\rangle $ is a first
excited state and $\theta \ll 1$), with the same $n$ as $|x_{n}\rangle $.
(iii) Rotate $|\psi (0)\rangle _{0}$ to $|\psi (0)\rangle =|g_{n,n\pm
1}\rangle +\theta ^{\prime }|e_{n,n\pm 1}\rangle $, a state that includes
contributions from $n\pm 1$ as well. (iv) Implement $U_{\mathrm{p}}(t)=\exp
(-iH_{\mathrm{p}}t)$ on $|\psi (0)\rangle $. (v) Measure. Repeat steps
(i)-(v) while increasing $t$ in step (v). We describe each of these steps in
detail, starting for simplicity from step (iv).

\textit{Step (iv): Implementation of }$\exp (-iH_{\mathrm{p}}t)$.--- In NMR
one can only control $f_{l}^{x}$ (or $f_{l}^{y}$) directly, while all $
\omega _{l},J_{l}$ are always on \cite{Cory:00}. Also, $J_{l}$ usually is
positive. A powerful method that allows us to deal with such constraints
(that are not unique to NMR) is \emph{recoupling} (e.g., \cite{Leung:00}).
The idea is based on elementary angular momentum theory. We define $C_{A}^{\varphi }\circ e^{i\theta B}\equiv e^{i\varphi
A}e^{i\theta B}e^{-i\varphi A}$, where $A,B$ are generators of $su(2)$
(e.g., two Pauli matrices), and/or $\{A,B\}=0$ while $A^{2}=\mathbf{1}$.
This \emph{recoupling sequence} can be interpreted as the application of
time-reversed pulses ($e^{\pm i\varphi A}$) before and after periods of free
evolution $e^{i\theta B}$. Special cases of interest are (i) $C_{A}^{\pi
/2}\circ e^{i\theta B}=e^{-i\theta B}$, (ii) $C_{A}^{\pi /4}\circ e^{i\theta
B}=e^{i\theta (iBA)}$. Thus, to obtain evolution under $\frac{\omega _{l}}{2}
\sigma _{l}^{z}$ we apply the (unoptimized) recoupling sequence $\exp (-
\frac{i\omega _{l}}{2}\sigma _{l}^{z}t)=(e^{-iH_{\mathrm{NMR}
}t/4}T_{l}e^{-iH_{\mathrm{NMR}}t/4}T_{l}^{\prime })^{2}$, where $
T_{l}=\otimes _{j\neq l}\sigma _{j}^{x}$, $T_{l}^{\prime }=\otimes _{j\neq
l}^{\prime }\sigma _{j}^{x}$ where the prime indicates that $j$ is even
(odd) if $l$ is even (odd). This takes $3N$ pulses. Fig.~1(a) illustrates an
optimized circuit for $N=2$. Similarly, we can evolve under any term $\sigma
_{j}^{z}\sigma _{j+1}^{z}$ using $\sim 7N$ recoupling steps.

\begin{figure}[tbp]
\includegraphics[height=12cm,angle=270]{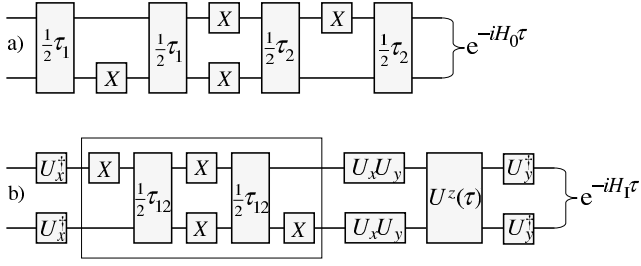} \vspace{-5cm}
\caption{Quantum circuits to simulate $e^{-iH_{0}\protect\tau }$ (a) and $
e^{-iH_{I}\protect\tau }$ (b) for the two qubit case. Time flows from left
to right. $X\equiv \protect\sigma ^{x}$. The recoupling procedure yielding $
U^{z}(\protect\tau )=\exp (-iJ_{1}\protect\tau _{12}\protect\sigma _{1}^{z}
\protect\sigma _{2}^{z})$ is in the box in (b), and is repeated without detail.
We set $\protect\omega _{i}\protect\tau _{i}=\protect\varepsilon _{i}\protect
\tau $ ($i=1,2$) and $2J_{1}\protect\tau _{12}=\left\vert V_{12}\right\vert 
\protect\tau $. Rectangular boxes connecting two qubits denote evolution
under $H_{\text{\textrm{NMR}}}$ for the indicated time. }
\label{fig1}
\end{figure}

Next, we need to show how to simulate \emph{long-range} interactions using $
H_{\mathrm{NMR}}$ and $F$. The set $\{X_{lm}\equiv \frac{1}{2}\left( \sigma
_{l}^{x}\sigma _{m}^{x}+\sigma _{l}^{y}\sigma _{m}^{y}\right) ,Y_{lm}\equiv 
\frac{1}{2}\left( \sigma _{l}^{y}\sigma _{m}^{x}-\sigma _{l}^{x}\sigma
_{m}^{y}\right) ,Z_{lm}\equiv \frac{1}{2}\left( \sigma _{l}^{z}-\sigma
_{m}^{z}\right) \}$ forms an $su(2)$ algebra, and commutes with $\sigma
_{m}^{z}+\sigma _{l}^{z}$ for any $l,m$ \cite{WuLidar:01}. Thus $
C_{X_{ll+1}}^{\pi /2}\circ Z_{l,l+1}=-Z_{l,l+1}$, while $C_{X_{ll+1}}^{\pi
/2}\circ (\sigma _{l}^{z}+\sigma _{l+1}^{z})=(\sigma _{l}^{z}+\sigma
_{l+1}^{z})$. Adding yields $C_{X_{ll+1}}^{\pi /2}\circ \left( \sigma
_{l-1}^{z}\sigma _{l}^{z}\right) =\sigma _{l-1}^{z}\sigma _{l+1}^{z}$, so
that $C_{X_{ll+1}}^{\pi /2}\circ e^{i\theta \sigma _{l-1}^{z}\sigma
_{l}^{z}}=e^{i\theta \sigma _{l-1}^{z}\sigma _{l+1}^{z}}$. Thus $
C_{X_{ll+1}}^{\pi /2}$ acts as a nearest-neighbor exchange operator. In
order to implement $C_{X_{ll+1}}^{\pi /2}$ using $H_{\mathrm{NMR}}$ and $F$
note that: 
\begin{equation*}
e^{-i\frac{\pi }{2}X_{ll+1}}=C_{\sigma _{l}^{x}+\sigma _{l+1}^{x}}^{-\pi
/4}\circ e^{{-i\frac{\pi }{4}\sigma _{l}^{z}\sigma _{l+1}^{z}}}\,C_{\sigma
_{l}^{y}+\sigma _{l+1}^{y}}^{\pi /4}\circ e^{{-i\frac{\pi }{4}\sigma
_{l}^{z}\sigma _{l+1}^{z}}}.
\end{equation*}
It is simple to check that to create all possible couplings $\sigma
_{l}^{z}\sigma _{m}^{z}$ in this manner requires $O(N^{3})$ steps. This
procedure allows us to use the short-range NMR Hamiltonian to simulate $
J_{l}\sigma _{l}^{z}\sigma _{m}^{z}$ with $|l-m|$ arbitrary. Let us now show how to turn this into a
simulation of $H_{I}\equiv \frac{1}{2}\sum_{l>m=1}^{N}V_{ml}(\sigma
_{m}^{x}\sigma _{l}^{x}+\sigma _{m}^{y}\sigma _{l}^{y})$. Suppose that $H_{
\mathrm{p}}$ evolves for time $\tau $. We can turn on $-J_{l}\sigma
_{l}^{z}\sigma _{m}^{z}$ for a time $\tau _{ml}$ such that $2J_{l}\tau
_{ml}=|V_{ml}|\tau $ (for a BCS\ Hamiltonian $V_{ml}<0$). Doing this for all
couplings separately (in series) shows that the evolution operator $
U^{z}(\tau )=\exp (-\frac{i}{2}\sum_{l>m}V_{lm}\sigma _{l}^{z}\sigma
_{m}^{z}\tau )$ is obtained using the same $O(N^{3})$ steps. By adjusting
single-qubit operation times, we can implement $U^{\alpha }=\exp (i\frac{\pi 
}{4}\sum_{l}\sigma _{l}^{\alpha })$, to yield: $\exp (-iH_{I}\tau )=\left(
U^{x\dagger }U^{z}(\tau )U^{x}\right) \left( U^{y}U^{z}(\tau )U^{y\dagger
}\right) $, using $O(N^{3})$ steps. However, $H_{\mathrm{p}}$ also contains
the term $H_{0}\equiv $$\sum_{l=1}^{N}\frac{\varepsilon _{l}}{2}\sigma
_{l}^{z}$, which does not commute with $H_{I}$. Clearly, by turning on
single qubit NMR $\sigma _{l}^{z}$ terms for times $\tau _{l}$ so that $
\omega _{l}\tau _{l}=\varepsilon _{l}\tau $, we can simulate $H_{0}$
directly using $N$ steps. The non-commutativity implies that we need a
short-time approximation in order to simulate the full $U_{\mathrm{p}}(\tau)=\exp (-iH_{\mathrm{p}}\tau )$:\ 
\begin{equation}
U_{\mathrm{p}}(\tau )=e^{-iH_{0}\tau }e^{-iH_{I}\tau }+O(\tau ^{2}).
\label{eq:BCSE}
\end{equation}
When the additional recoupling steps needed to turn off unwanted
interactions (which we ignored above) are taken into account, using the
method of \cite{Leung:00}, we find that $U_{\mathrm{p}}(\tau )$ requires a
total of $\allowbreak s(N)=-\frac{4}{3}N^{2}+\frac{32}{3}N-\frac{47}{3}N^{3}+
\frac{28}{3}N^{4}$ steps. This result may be
improved somewhat if parallel operations are allowed. E.g., in Fig.~1 we show
optimized circuits implementing $e^{-iH_{0}\tau }$ and $e^{-iH_{I}\tau }$
for $N=2$ qubits. If $H_{\mathrm{NMR}}$ contains beyond-nearest-neighbor
interactions then at most $O(N^{5})$ steps are needed. The effect of the $
O(\tau ^{2})$ errors in quantum algorithms due to the short-time
approximation has been analyzed, e.g., in \cite{Dodd:01Wocjan:01}.
By concatenating short-time evolution segments one can then obtain the
finite time ($k\tau =t$) evolution operator $U_{\mathrm{p}}(t)\approx (U_{
\mathrm{p}}(\tau ))^{k}$ \cite{Abrams:99}, in a total of $k\,s(N)$ steps.

\textit{Step (ii):\ Adiabatic Evolution}.---\textit{\ }Let $2\Delta $ be the
gap between the ground and the first excited states, and let $0\leq c(t)\leq
1$, $c(0)=0$, $c(T)=1$, be a slowly varying function, i.e., $2\pi /T\ll
2\Delta $ (e.g., $c(t)=t/T$). Consider the time-ordered evolution $U_{
\mathrm{ad}}(t)=\mathcal{T}\exp (-i\int_{0}^{t}H(s)ds)$ under a
time-dependent Hamiltonian $H(t)=H_{0}+c(t)H_{I}$. For sufficiently small $
\tau $ this factors into a product 
\begin{equation}
U_{\mathrm{ad}}(k\tau )\approx e^{-iH(k\tau )\tau }\cdots e^{-iH(2\tau )\tau
}e^{-iH(\tau )\tau }+O(\tau ^{2}),  \label{eq:ad}
\end{equation}
where $\exp (-iH(j\tau )\tau )\approx \exp (-iH_{0}\tau )\exp (-ic(j\tau)H_{I}\tau )$ ($j=1,...,k$), and now we choose times $\tau _{ml}(j)$ (for
turning on $-J_{l}\sigma _{l}^{z}\sigma _{m}^{z}$) such that $2J_{l}\tau_{ml}()=|V_{ml}|\tau \,c(j\tau )$. Since $c(t)$ is slow, $U_{\mathrm{ad}
}(k\tau )$ will represent an \emph{adiabatic evolution}. The adiabatic
theorem then ensures that the system will be in an eigenstate of $H_{\mathrm{
p}}=H(T)$ at $T=k\tau $, provided the initial state is in an eigenstate of $
H_{0}$. Moreover, this will be a ground state $|g_{n}\rangle $ of $H_{
\mathrm{p}}$ (a state with fixed $n$) if the initial state is the ground
state of $H_{0}$ (a computational basis state $|x_{n}\rangle $) \cite{comment}.
In order to probe the low-lying
spectrum we may slightly relax the adiabatic condition $\pi /T\ll \Delta $,
or $k\gg \pi /(\tau \Delta )$. This can be defined in terms of the adiabatic
expansion where the first order constraint is the usual adiabatic
assumption. Here we only wish to satisfy the second order condition
\cite{Wu:89}. Then we obtain a state $|\psi (0)\rangle _{0}\approx
|g_{n}\rangle 
+\theta |e_{n}\rangle $ which contains a small ($\theta \ll 1$) component $
|e_{n}\rangle $ of some of the low-lying excited states of $H_{\mathrm{p}}$
(with the same $n$).

\textit{Steps (iii),(v):\ Measuring the Spectrum}.--- In NMR one
measures the free-induction-decay (FID) signal, given by $V_{\alpha
}(t)\propto \mathrm{Tr}(\rho (t)\sigma _{\alpha }^{-})$, where $\rho (t)$ is
the system density matrix and $\alpha $ is the index of the measured spin
(qubit) \cite{Cory:00}. To probe states with different $n$, we rotate
to $|\psi (0)\rangle =e^{-i\omega \sigma _{\alpha }^{y}}|\psi (0)\rangle
_{0}\approx |g_{n,n\pm 1}\rangle +\theta ^{\prime }|e_{n,n\pm 1}\rangle $,
where $\theta ^{\prime },\omega \ll 1$, a state that includes contributions
from $n\pm 1$ as well [\textit{step (iii)}]. This is simple to do using the
method of step (iv). Combining steps (ii)-(iv), we have $\rho (t)=U_{\mathrm{
p}}(t)|\psi (0)\rangle \langle \psi (0)|U_{\mathrm{p}}^{\dagger }(t)$. To
relate $V_{\alpha }(t)$ to the spectrum of the pairing
Hamiltonian we introduce an appropriate basis. A complete set of
conserved quantum numbers are the number of Cooper pairs $n$ (= the number
of $1$'s in a computational basis state, lowered by $\sigma _{\alpha }^{-}$), the energy $E_{n,i}$ for fixed $n$, and a state degeneracy index $\beta
_{i}$. Thus our basis states are labeled by $|n,i,\beta _{i}\rangle $ and $
\rho (t)$ can be expanded as $\sum B_{n,i,\beta _{i}}B_{m,j,\beta
_{j}}^{\ast }|n,i,\beta _{i}\rangle e^{i(E_{m,j}-E_{n,i})t}\langle m,j,\beta
_{j}|$ with $|\psi (0)\rangle =\sum_{n,i,\beta _{i}}B_{n,i,\beta
_{i}}|n,i,\beta _{i}\rangle $. We have 
\begin{equation}
V_{\alpha }(t)\propto \sum_{m,n}\sum_{i,j}C_{m,j;n,i}^{(\alpha)}e^{i(E_{m,j}-E_{n,i})t},  \label{eq:Vs}
\end{equation}
where $C_{m,j;n,i}^{(\alpha )}\equiv \sum_{\beta _{i}\beta _{j}}B_{n,i,\beta
_{i}}B_{m,j,\beta _{j}}^{\ast }\langle m,j,\beta _{j}|\sigma _{\alpha
}^{-}|n,i,\beta _{i}\rangle $ $\propto $$\delta _{m,n-1}$. Fourier
transforming, we obtain the energy spectrum $S(\omega )=\sum_{n,i,j}\tilde{C}
_{n-1,j;n,i}^{(\alpha )}\delta (\omega -(E_{n-1,j}-E_{n,i}))$, with the gap
defined as $2\Delta _{n}\equiv E_{n,1}-E_{n,0}$. Ideally, $\Delta _{n}$ can
be found from a few runs with different initial $n$. There are two
complications in practice: (i) Finding $\Delta _{n}$ in this manner depends
on the coefficients $\tilde{C}_{n-1,j;n,i}^{(\alpha )}$ not vanishing. By
measuring all qubits $\alpha $, it is likely that sufficiently many non-zero
coefficients will be available. (ii) The sharpness of the $\delta $
functions depends on how densely the signal $V_{\alpha }(t)$ is sampled. To
resolve the gap, we will need to sample with a resolution $\Delta \omega
=2\pi /T<\Delta _{n}$. Recall that $H_{\mathrm{BCS}}$ conserves $n$. Thus
the number of $\tau $-intervals required for fixed $n$ is $k(n)\gg \pi
/(\tau \Delta _{n})$, which is just the adiabatic condition again. A total
of $\frac{1}{2}k(n)^{2}$ elementary evolutions steps, each simulating
evolution under $H_{\mathrm{p}}$ for length $\tau $, will thus be needed to
simulate $\{U_{\mathrm{p}}(k\tau )\}_{k=1}^{T/\tau }$, and each
such step takes $s(N)$ logic gates. The longest single run takes $k(n)s(N)$
steps, while $\frac{1}{2}k(n)^{2}s(N)$ is the total run-time of the
algorithm. if the algorithm is to succeed in the absence of error
correction, then we must have $k(n)s(N)<T_{2}/\tau _{\mathrm{\log ic}}$, the
ratio of decoherence to logic gate time. For NMR, $T_{2}/\tau _{\mathrm{\log
ic}}$ can be $\sim 10^{5}$. To estimate $k(n)$ we need $\tau $ and $\Delta
_{n}$. The gap can be estimated experimentally, for nuclear and BCS systems
using material dependent parameters \cite{Mahan:bookRing:book,Ralph:97}. 
Recall that $
\tau $ is related to the short-time approximation which allowed us to
neglect commutator terms in the expansion of $U_{\mathrm{ad}}(t)$. Since $
e^{\left( A+B\right) \tau }\approx e^{A\tau }e^{B\tau }e^{-\frac{1}{2}
[A,B]\tau ^{2}}$, we need to estimate when $|[A,B]\tau |\ll \min (|A|,|B|)$.
To obtain a rough estimate we consider a reduced BCS model 
\cite{Dukelsky:99}: $
V_{ml}\equiv -V<0$, $\varepsilon _{l}=\varepsilon _{0}+ld$. In the BCS case
the level spacing $d\ll V$, but $\varepsilon _{0}\gg V$. Letting $
A=\varepsilon _{l}\sigma _{l}^{z}$, $B=VX_{lm}$, we have $|[A,B]|=|V\left(
\varepsilon _{l}-\varepsilon _{m}\right) Y_{lm}|>Vd$, while $\min
(|A|,|B|)=V $. Thus the short-time approximation is valid when $\tau \ll 1/d.
$ Using $k(n)\gg \pi /(\tau \Delta _{n})$ and $s(N)\approx 9N^{4}$ we thus
have $k(n)\,s(N)\gg 30\frac{d}{\Delta _{n}}N^{4}$. In the BCS case $d/\Delta
_{n}\ll 1$. Assuming $d/\Delta _{n}=0.1$ we find $k(n)\,s(10)\gg 3\times
10^{4}$, so that a simulation with $N \le 10$ qubits seems to be within the
reach of present day NMR simulations \cite{Cory:00}.

In order to illustrate the algorithm, consider a simple example, the circuit
for which is given in Fig.~1. When $N=2$ the computational basis states are: $
\{\left\vert 00\right\rangle ,\left\vert 01\right\rangle ,\left\vert
10\right\rangle ,\left\vert 11\right\rangle \}$, with $n=0,1,1,2$ Cooper
pairs, respectively. Diagonalizing $H_{\mathrm{p}}$ yields the energy
spectrum:\ $\{E_{n}\}=\{E_{0}=-(\varepsilon _{1}+\epsilon _{2})/2,E_{1}^{\pm
}=\pm \sqrt{\epsilon ^{2}+V^{2}},E_{2}=(\varepsilon _{1}+\epsilon _{2})/2\}$. Steps (ii)-(v) of the algorithm can be carried out analytically. Fourier
transforming the FID signal yields four spectral lines from which, e.g.,
the $n=1$ gap can be found as $2\Delta _{1}=E_{1}^{+}-E_{1}^{-}$.

\textit{Conclusions.---} We have proposed an efficient algorithm for finding
the low-lying spectrum of pairing models with arbitrary long-range
interactions, such as the BCS Hamiltonian. This establishes a link between
quantum computers (QCs) of the next generation (10-100 qubits) and
outstanding problems in finite-system quantum physics, such as the
applicability of the BCS model to mesoscopic solid-state and nuclear
systems. It would be interesting to implement the algorithm using current
NMR-QC know-how, thus extending the experimental repertoire of QC physics
simulations \cite{Somaroo:99Tseng:00}.

D.A.L. gratefully acknowledges financial support from PREA, NSERC, and the
Connaught Fund.

\end{document}